\documentclass[a4paper,11pt]{article}
\usepackage{pos}

\def\apj{ApJ}                 
\def\aap{A\&A}                
\def\mnras{MNRAS}             

\def\ssr{Space Science Reviews}

\title{Deep observations of Kepler's SNR with H.E.S.S.}
 \ShortTitle{Kepler's SNR with H.E.S.S.}

\author*[a]{Dmitry Prokhorov}
\author[a]{Jacco Vink}
\author[a]{Rachel Simoni}
\author[c]{Nukri Komin}
\author[d]{Stefan Funk}
\author[e]{Denys Malyshev}
\author[d]{Lars Mohrmann}
\author[f]{Stefan Ohm}
\author[e]{Gerd P{\"u}hlhofer}
\author[h]{Heinrich J. V{\"o}lk}

\affiliation[a]{GRAPPA, Anton Pannekoek Institute for Astronomy, University of Amsterdam, Science Park 904, 1098 XH Amsterdam, The Netherlands}

\affiliation[c]{School of Physics, University of the Witwatersrand, 1 Jan Smuts Avenue, Braamfontein, Johannesburg, 2050 South Africa}

\affiliation[d]{Friedrich-Alexander-Universit{\"a}t Erlangen-N{\"u}rnberg, Erlangen Centre for Astroparticle Physics, Erwin-Rommel-Str. 1, D 91058 Erlangen, Germany}

\affiliation[e]{Institut f{\"u}r Astronomie und Astrophysik, Universit{\"a}t T{\"u}bingen, Sand 1, D 72076 T{\"u}bingen, Germany}

\affiliation[f]{DESY, D-15738 Zeuthen, Germany}

\affiliation[h]{Max-Planck-Institut für Kernphysik, P.O. Box 103980, D 69029 Heidelberg, Germany}

\forColl{H.E.S.S.} 

\emailAdd{d.prokhorov@uva.nl}
\emailAdd{j.vink@uva.nl}
\emailAdd{r.c.simoni@uva.nl}

\abstract{Kepler's supernova remnant (SNR) which is produced by the most recent naked-eye supernova in our Galaxy is one of the best studied SNRs, but its gamma-ray detection has eluded us so far. Observations with modern imaging atmospheric Cherenkov telescopes (IACT) have enlarged the knowledge about nearby SNRs with ages younger than 500 years by establishing Cassiopeia A and Tycho's SNRs as very high energy (VHE) gamma-ray sources and setting a lower limit on the distance to Kepler's SNR. This SNR is significantly more distant than the other two and expected to be one of the faintest gamma-ray sources within reach of the IACT arrays of this generation. We report strong evidence for a VHE signal from Kepler's SNR based on deep observations of the High Energy Stereoscopic System (H.E.S.S.) with an exposure of 152 hours, including 122 hours accumulated in 2017-2020. We further discuss implications of this result for cosmic-ray acceleration in young SNRs.}

\FullConference{37$^{\rm{th}}$ International Cosmic Ray Conference (ICRC 2021)\\
		July 12th -- 23rd, 2021\\
		Online -- Berlin, Germany}

\begin{document}
\maketitle

\section{Introduction}

It is believed that the supernova remnants (SNRs) can convert $\gtrsim$10\% of their SN explosion energy into the energy of relativistic atomic nuclei (the so-called hadronic cosmic rays), accelerated at their shock fronts.  Given the supernova rate and explosion energies, SNRs are considered to be the main sources of cosmic rays with energies up to $3\times10^{15}$ eV.   

The remnant of the youngest naked-eye supernova SN 1604, better known as Kepler’s SNR, is among the best studied SNRs across the whole electromagnetic spectrum \cite[for a review,][]{Vink2017}. Both the historical light curve and the X-ray spectra show that SN 1604 belongs to the class of normal Type Ia SNe, which explode with energies of $10^{51}$ erg. The dynamics of the SNR has been well studied, indicating shock velocities of 2000-5300$d_5$ km s$^{-1}$, with $d_5$ the distance in units of 5 kpc. The distance was long uncertain, but a reassessment of the historical light curve and new proper motion/spectral studies in the optical band showed that $d=5\pm1$ kpc \cite{sankrit16, Ruiz2017}. The large range in shock velocities, as well as infrared (IR) and optical studies indicate that the SNR is interacting with a dense and clumpy medium in the northwest and along the central bar, which, given the height above the Galactic plane of 600$d_5$ pc likely originates from the progenitor system itself \cite{Vink2017}. 

The high overall shock velocities are ideal for accelerating cosmic rays. And indeed, the X-ray continuum reveals thin regions of synchrotron emission demarcating the shock in Kepler's SNR. The presence of these filaments show that electrons are accelerated to $>10^{13}$ eV and the filament widths show that the magnetic fields must be larger than 100 $\mu$G \cite{helder12}. The width of these filaments are only comparable to those at shock regions of Cassiopeia A and Tycho's SNR \cite{helder12}, which all provide the highest levels of magnetic field amplification by cosmic-ray streaming. Hence, these shocks should efficiently accelerated particles. 

Kepler’s SNR is the only historical SNR which is absent from the list of young SNRs detected at very high energies (VHE, > 100 GeV). The High Energy Stereoscopic System (H.E.S.S.) Cherenkov telescopes did observe Kepler’s SNR in the past (in 2004-2005) \cite{kepler2008}. The previous observations did not result in a detection of the remnant, but in a flux upper limit. This is partially due to the relatively short exposure time compared to other SNRs, but partially also due to the fact that Kepler’s SNR is more distant than the other historical SNRs. 

We present the results of deep observations of Kepler's SNR performed with H.E.S.S. based on 152 hours of the observations,
including 122 hours of observations in 2017-2020. The exposure time of these observations exceeds that of the previously reported observations \cite{kepler2008} by more than 10 times. We report strong evidence for a VHE signal from Kepler’s SNR in the deep H.E.S.S. observations at a statistical level of 4.6$\sigma$. It indicates that Kepler's SNR is a VHE gamma-ray source and confirms the presence of particles accelerated to TeV energies. The strong evidence for the VHE signal from Kepler's SNR is supported by the presence of a tentative source in the \textit{Fermi}-LAT data in the direction of Kepler's SNR as shown by our analysis. The combination of high-energy and VHE gamma-ray results is also important for determining whether the gamma-ray emission is dominated by hadronic or leptonic emission processes.  

\section{Observations and results}

\subsection{H.E.S.S. data}

H.E.S.S. is a system of five Imaging Cherenkov Telescopes, located in the Khomas Highland of Namibia at an altitude of 1800 m. Located in the southern hemisphere it is well-suited for VHE observations of Kepler's SNR. In 2017-2020, the H.E.S.S. array consisted of four upgraded 12 m-diameter telescopes placed in a square with 120 m sides and one 28 m-diameter telescope (H.E.S.S. phase II array) in the center of the array. H.E.S.S. employs the stereoscopic imaging atmospheric Cherenkov technique. Dedicated observations of Kepler's SNR with
H.E.S.S. were performed in wobble mode with offsets by
0.7$^{\circ}$ from Kepler’s SNR, allowing a simultaneous measurement of the background in the same field of view. Observations of Kepler's SNR were conducted during the May-October visibility window.

\begin{figure*}
\centering
    \includegraphics[angle=0, width=.7\textwidth]{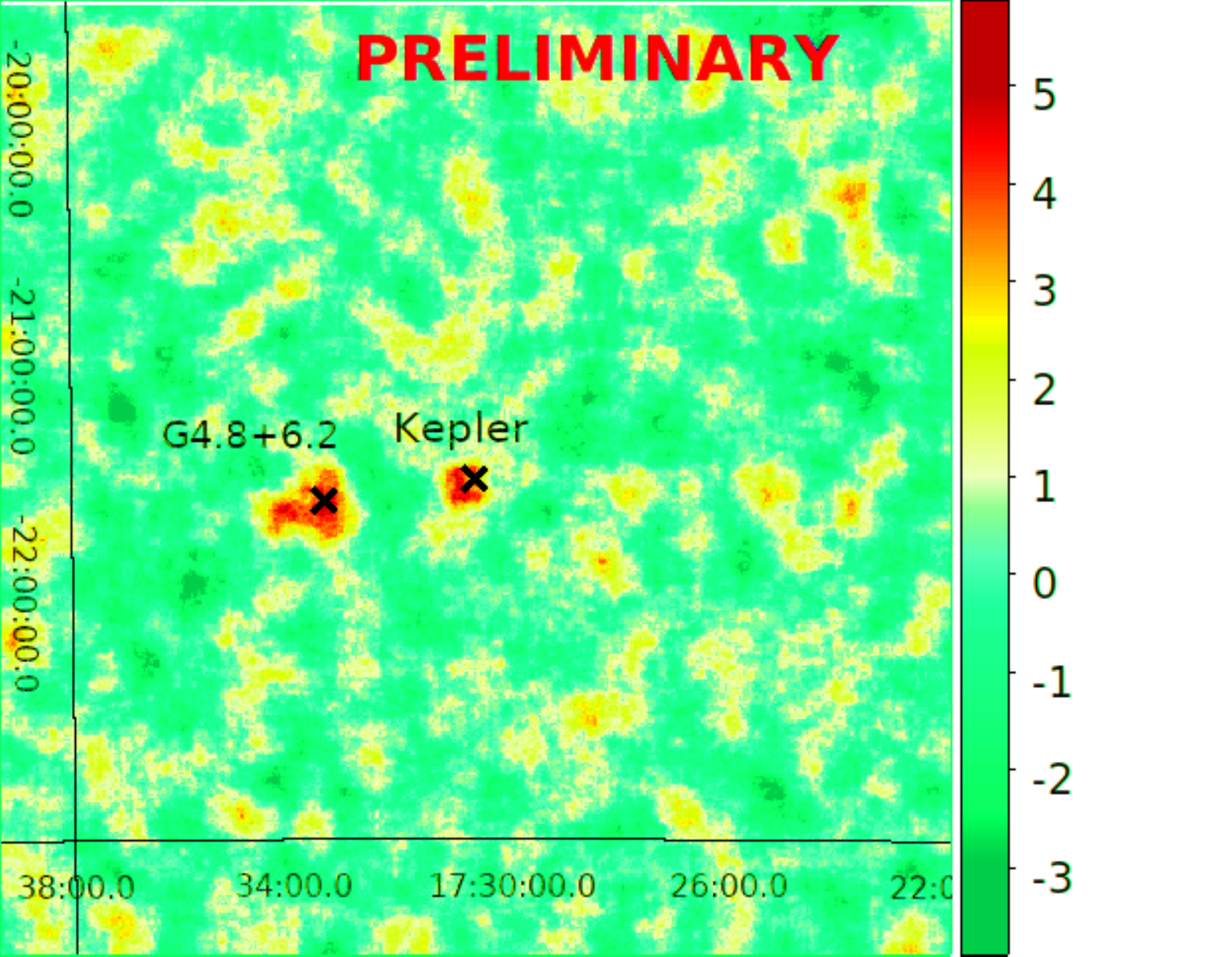}
  \caption{H.E.S.S. gamma-ray significance map
of Kepler's SNR using an integration radius of 0.1 degree. 
The positions of Kepler's SNR and SNR G4.8+6.2 are shown with crosses.} 
\label{F1}
\end{figure*}

A standard data quality selection procedure was used to identify observations with the satisfactory hardware state of the cameras and good atmospheric conditions. The data were analyzed using the Model Analysis \cite[][]{Modelplus} and the analysis configuration, which requires a minimum of 60 photo-electrons per image and considers events with an estimated direction reconstruction uncertainty of less than 0.1$^{\circ}$. The results were cross-checked with the Image Pixel-wise fit for Atmospheric Cherenkov Telescope (ImPACT) analysis \cite{ImPACT2014}.

\begin{figure*}
\centering
    \includegraphics[angle=0, width=.7\textwidth]{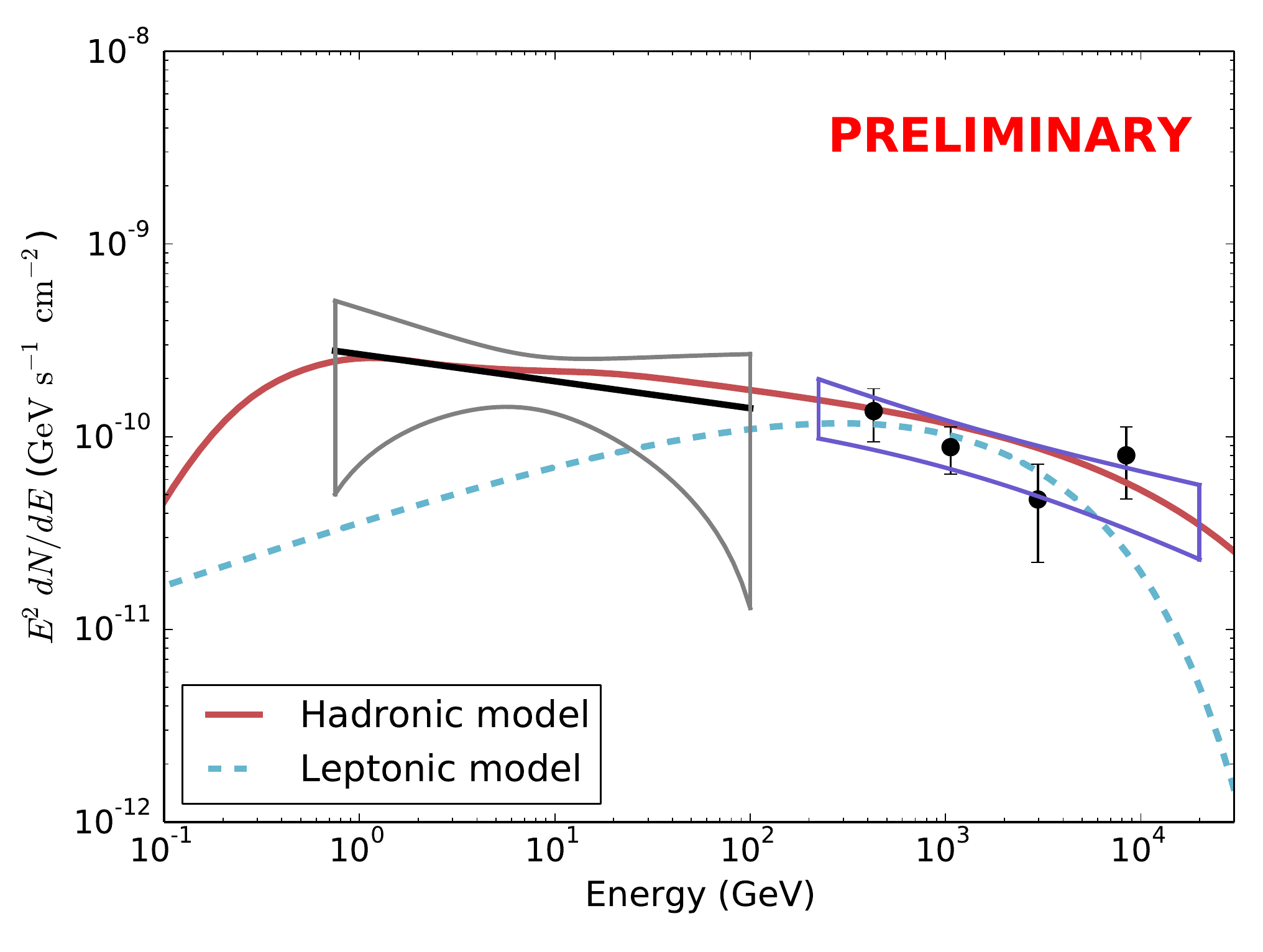}
  \caption{SED of Kepler's SNR. \textit{Fermi-LAT} ($<100$ GeV) and H.E.S.S. data ($>226$ GeV) 
points along with the hadronic and leptonic models.} 
\label{F2}
\end{figure*}

The background subtraction was performed using the standard algorithms used in H.E.S.S. - the ring background method (for sky maps) and the reflected-region background method (for spectral measurements), see \cite{Berge2007}. The region around another potential VHE gamma-ray source (SNR G4.8+6.2) in the field of view was excluded from background estimation. At the nominal position of Kepler's SNR an excess of 178 gamma rays above the background was detected by us with a statistical significance of 4.6$\sigma$. The energy spectrum was derived using a forward-folding technique. The analysis energy threshold for this data set is 226 GeV. The H.E.S.S. significance map is in Figure \ref{F1}.
The spectral analysis resulted in points shown in Figure \ref{F2}.

\subsection{\textit{Fermi}-LAT data}

The Large Area Telescope \cite{Atwood2009} is a pair-conversion telescope, covering the energy range from about 20 MeV to more than 300 GeV, onboard the Fermi Gamma-ray Space Telescope.

For the data analysis, the Fermitools package and P8R3\_SOURCE\_V2 instrument response functions were used. For this analysis, LAT gamma-ray events with reconstructed energies between 750 MeV and 300 GeV and accumulated from 2008 August 4 to 2019 May 16 were selected, but those with a zenith angle larger than $90^{\circ}$ were excluded. Standard quality cuts (DATA\_QUAL$>0$ \&\& LAT\_CONFIG==1) were applied. To model the sources within the ROI, sources from the 4FGL catalog were included. The normalization and photon index of Kepler's SNR, the normalizations of Galactic and isotropic diffuse
sources, \texttt{gll\_iem\_07.fits} and \texttt{iso\_P8R3\_SOURCE\_V2\_v1.txt}, and the normalizations of 4FGL gamma-ray sources within 3$^{\circ}$ from Kepler’s SNR were allowed to vary, while the normalizations of other 4FGL sources were held fixed. 

Our analysis resulted in a test-statistic \cite[TS;][]{Mattox} value for Kepler’s SNR of 16.8, which corresponds to a 4 sigma significance. The \textit{Fermi}-LAT TS map reveals similarities with the H.E.S.S. significance map indicating the presence of both the SNRs, Kepler's SNR and G4.8+6.2. 
The spectral butterfly plot obtained on the basis of the \textit{Fermi}-LAT data is in Figure \ref{F2}.

The GeV counterparts of Kepler's SNR and SNR G4.8+6.2, if considered together with the corresponding VHE excesses seen with H.E.S.S., provide strong support for identification
of Kepler's SNR in the GeV and VHE gamma-ray bands\footnote{We took the results presented in
this section into account when we made the decision to perform H.E.S.S. observations of
Kepler's SNR in 2020. During the preparation of the H.E.S.S. and \textit{Fermi}-LAT results reported in this paper, we became aware of the results of \cite{Xiang2021}, who derived similar evidence for the GeV excess toward Kepler's SNR from \textit{Fermi}-LAT data.}.

The gamma-ray excesses at GeV and TeV energies at the location of SNR G4.8+6.2 come as a surprise. This SNR candidate is not well studied, so its physical properties are not well known. In the radio band, SNR G4.8+6.2 has a shell-like morphology and an angular extent of 18$^{\prime}$ at 1.4 GHz (the NRAO VLA Sky Survey). At 2.3 GHz it appears highly polarized with an almost constant orientation of the polarization vectors across the source and with the mean fraction of polarized emission of up to 25\% \cite{Duncan1997}. Young SNRs, such as Kepler's SNR, have a much smaller fractional polarization. Given that G4.8+6.2 comes out of a blind search for sources using observations targeting Kepler's SNR, the a-posteriori significance of of Kepler's SNR is higher. More details will be published in a forthcoming paper.

\section{Interpretation}

Given the gamma-ray spectral properties of Tycho's SNR and Cassiopeia A, their gamma-ray emissions are likely of hadronic origin \cite{Morlino2012, casa2017}. The leptonic scenario, in which inverse Compton mechanism dominates the VHE emission, is still a viable scenario for Tycho's SNR, but only for its TeV emission, and under the assumption that its GeV gamma-ray emission is due to hadronic interactions.

To construct a characteristic spectral energy distribution (SED) of Kepler's SNR, we used the H.E.S.S. and \textit{Fermi}-LAT data points reported in this paper and additionally included the archival data points in the radio and X-ray bands. We used a package, \texttt{Naima} \cite{Zabalza2015}, to model these data. More information about the SED modeling will be given in the forthcoming journal paper \cite{Kepler2022}. We show both the hadronic and leptonic models in Figure \ref{F2}. The values of physical parameters used for the hadronic model are: the SN Ia explosion energy, $10^{51}$ erg; the cosmic-ray hadron energy of 10\% of the SN Ia explosion energy; the gas target particle density, 0.5 cm$^{-3}$, the cosmic-ray proton spectral index, 2.2; and the exponential cut-off in the cosmic-ray proton spectrum at 300 TeV. 

The values of the physical parameters used for the leptonic model are: the cosmic-ray electron energy of 0.15\% of the SN Ia explosion energy; the cosmic-ray electron spectral index, 2.3; the magnetic field strength, 80 $\mu$G; the exponential cut-off energy of the cosmic-ray electron spectrum is 11 TeV. To compute the inverse Compton gamma-ray component, we include three soft photon fields, the cosmic microwave background (CMB), the infrared photon field emitted by dust in the SNR, and the Galactic infrared photon field. We found that the derived LAT photon index of $\Gamma=2.12\pm0.32$ is softer than the value expected  for the leptonic model, $\Gamma=1+\alpha=1.71$, with $\alpha=0.71$ the radio spectral index. But the two values are marginally compatible within error bars. However, the magnetic field strength of 80~${\rm \mu G}$ required by the leptonic model is smaller than inferred from the non-thermal X-ray filaments.

The hadronic model for emission at both GeV and TeV energies is thus preferred on physical grounds. It is indicative that the SED and preferred radiation models for Tycho's and Kepler's SNRs are rather similar. Since Kepler's SNR is fainter than Tycho's SNR and the associated flux uncertainties are larger on account of its larger distance, a situation similar to that for Tycho's SNR, in that the
inverse Compton mechanism is viable only for the TeV emission, but not for the GeV emission, is possible. 

\section{Summary}

The H.E.S.S. observations with exposure time of 152 hours resulted in significant evidence for VHE gamma-ray emission from Kepler's SNR. This confirms the presence of particles at energies over 1 TeV accelerated in Kepler's SNR as previously established in the X-ray band. Given that the other near SNR younger than 500 years old, Cassiopeia A and Tycho, have previously been revealed as VHE gamma-ray sources, our results obtained for Kepler's SNR support that the production of VHE emission is a general property of SNR with age of about 400 years. \textit{Fermi}-LAT observations provides a further support for identification of this gamma-ray signal with Kepler's SNR.  

\section{Acknowledgements}

The support of the Namibian authorities and of the University of Namibia in facilitating the construction and operation of H.E.S.S. is gratefully acknowledged, as is the support by the German Ministry for Education and Research (BMBF), the Max Planck Society, the German Research Foundation (DFG), the Helmholtz Association, the Alexander von Humboldt Foundation, the French Ministry of Higher Education, Research and Innovation, the Centre National de la Recherche Scientifique (CNRS/IN2P3 and CNRS/INSU), the Commissariat à l'énergie atomique et aux énergies alternatives (CEA), the U.K. Science and Technology Facilities Council (STFC), the Knut and Alice Wallenberg Foundation, the National Science Centre, Poland grant no. 2016/22/M/ST9/00382, the South African Department of Science and Technology and National Research Foundation, the University of Namibia, the National Commission on Research, Science \& Technology of Namibia (NCRST), the Austrian Federal Ministry of Education, Science and Research and the Austrian Science Fund (FWF), the Australian Research Council (ARC), the Japan Society for the Promotion of Science and by the University of Amsterdam.

We appreciate the excellent work of the technical support staff in Berlin, Zeuthen, Heidelberg, Palaiseau, Paris, Saclay, T\"{u}bingen and in Namibia in the construction and operation of the equipment. This work benefitted from services provided by the H.E.S.S. Virtual Organisation, supported by the national resource providers of the EGI Federation.

%
%
%
\vfill
\pagebreak
\section*{Full Authors List: H.E.S.S. Collaboration}
\scriptsize
\noindent
H.~Abdalla$^{1}$, 
F.~Aharonian$^{2,3,4}$, 
F.~Ait~Benkhali$^{3}$, 
E.O.~Ang\"uner$^{5}$, 
C.~Arcaro$^{6}$, 
C.~Armand$^{7}$, 
T.~Armstrong$^{8}$, 
H.~Ashkar$^{9}$, 
M.~Backes$^{1,6}$, 
V.~Baghmanyan$^{10}$
% ...

\medskip

\noindent
$^{1}$University of Namibia, Department of Physics, Private Bag 13301, Windhoek 10005, Namibia\\
$^{2}$Dublin Institute for Advanced Studies, 31 Fitzwilliam Place, Dublin 2, Ireland\\
$^{3}$Max-Planck-Institut f\"ur Kernphysik, P.O. Box 103980, D 69029 Heidelberg, Germany\\
$^{4}$High Energy Astrophysics Laboratory, RAU,  123 Hovsep Emin St  Yerevan 0051, Armenia\\
$^{5}$Aix Marseille Universit\'e, CNRS/IN2P3, CPPM, Marseille, France\\
$^{6}$Centre for Space Research, North-West University, Potchefstroom 2520, South Africa\\
$^{7}$Laboratoire d'Annecy de Physique des Particules, Univ. Grenoble Alpes, Univ. Savoie Mont Blanc, CNRS, LAPP, 74000 Annecy, France\\
$^{8}$University of Oxford, Department of Physics, Denys Wilkinson Building, Keble Road, Oxford OX1 3RH, UK\\
$^{9}$IRFU, CEA, Universit\'e Paris-Saclay, F-91191 Gif-sur-Yvette, France\\
$^{10}$Instytut Fizyki J\c{a}drowej PAN, ul. Radzikowskiego 152, 31-342 Krak{\'o}w, Poland\\
% ...


\begin{thebibliography}{99}
\bibitem[{{Vink}(2017)}]{Vink2017}
{Vink}, J., \textit{Supernova 1604, Kepler's Supernova, and its Remnant}, 2017, {Supernova 1604, Kepler's Supernova, and its Remnant}, ed. A.~W. Alsabti \& P.~Murdin (Cham, Switzerland: Springer International
  Publishing), 139--160
  
  \bibitem[{{Sankrit} {et~al.}(2016){Sankrit}, {Raymond}, {Blair}, {Long},
  {Williams}, {Borkowski}, {Patnaude}, \& {Reynolds}}]{sankrit16}
{Sankrit}, R., {Raymond}, J.~C., {Blair}, W.~P., {et~al.}, \textit{Second epoch Hubble space telescope observations of Kepler's supernova remnant: the proper motions of Balmer filaments}, 2016, \apj, 817, 36

\bibitem[{{Ruiz-Lapuente}(2017)}]{Ruiz2017}
{Ruiz-Lapuente}, P., \textit{The light curve and distance of the Kepler supernova: news from four centuries ago}, 2017, \apj, 842, 112

\bibitem[{{Helder} {et~al.}(2012{\natexlab{a}}){Helder}, {Vink}, {Bykov},
  {Ohira}, {Raymond}, \& {Terrier}}]{helder12}
{Helder}, E.~A., {Vink}, J., {Bykov}, A.~M., {et~al.}, \textit{Observational signatures of particle acceleration in supernova remnants}, 2012{\natexlab{a}}, \ssr,
  173, 369

\bibitem[{{Aharonian} {et~al.}(2008{\natexlab{a}}){Aharonian}, {Akhperjanian},
  {Barres de Almeida}, {Bazer-Bachi}, {Behera}, {Beilicke}, {Benbow}, {Berge},
  {Bernl{\"o}hr}, \& {Boisson}}]{kepler2008}
{Aharonian}, F., {Akhperjanian}, A.~G., {Barres de Almeida}, U., {et~al.}, \textit{HESS upper limits for Kepler's supernova remnant}, 2008{\natexlab{a}}, \aap, 488, 219  
  
  \bibitem[{{de Naurois} \& {Rolland}(2009)}]{Modelplus}
{de Naurois}, M. \& {Rolland}, L., \textit{A high performance likelihood reconstruction of $\gamma$-rays for imaging atmospheric Cherenkov telescopes}, 2009, Astroparticle Physics, 32, 231

\bibitem[{{Parsons} \& {Hinton}(2014)}]{ImPACT2014}
{Parsons}, R.~D. \& {Hinton}, J.~A., \textit{A Monte Carlo template based analysis for air-Cherenkov arrays}, 2014, Astroparticle Physics, 56, 26
  
\bibitem[{{Berge} {et~al.}(2007){Berge}, {Funk}, \& {Hinton}}]{Berge2007}
{Berge}, D., {Funk}, S., \& {Hinton}, J., \textit{Background modelling in very-high-energy $\gamma$-ray astronomy}, 2007, \aap, 466, 1219  

\bibitem[{{Atwood} {et~al.}(2009){Atwood}, {Abdo}, {Ackermann}, {Althouse},
  {Anderson}, {Axelsson}, {Baldini}, {Ballet}, {Band}, {Barbiellini},
  {Bartelt}, {Bastieri}, {Baughman}, {Bechtol}, {B{\'e}d{\'e}r{\`e}de},
  {Bellardi}, {Bellazzini}, {Berenji}, {Bignami}, {Bisello}, {Bissaldi}, \&
  {Blandford}}]{Atwood2009}
{Atwood}, W.~B., {Abdo}, A.~A., {Ackermann}, M., {et~al.}, \textit{The Large Area Telescope on the Fermi Gamma-Ray Space Telescope Mission}, 2009, \apj, 697, 1071
  
\bibitem[{{Mattox} {et~al.}(1996){Mattox}, {Bertsch}, {Chiang}, {Dingus},
  {Digel}, {Esposito}, {Fierro}, {Hartman}, {Hunter}, {Kanbach}, {Kniffen},
  {Lin}, {Macomb}, {Mayer-Hasselwander}, {Michelson}, {von Montigny},
  {Mukherjee}, {Nolan}, {Ramanamurthy}, {Schneid}, {Sreekumar}, {Thompson}, \&
  {Willis}}]{Mattox}
{Mattox}, J.~R., {Bertsch}, D.~L., {Chiang}, J., {et~al.}, \textit{The likelihood analysis of EGRET data}, 1996, \apj, 461, 396

\bibitem[{{Xiang} \& {Jiang}(2021)}]{Xiang2021}
{Xiang}, Y. \& {Jiang}, Z., \textit{Detection of GeV gamma-ray emission of Kepler's SNR With Fermi-LAT}, 2021, \apj, 908, 22

\bibitem[{{Duncan} {et~al.}(1997) {Duncan}, {Stewart}, {Haynes}, \& 
{Jones}}] {Duncan1997} 
{Duncan}, A. R., {Stewart}, R. T., {Haynes}, R. F. \& {Jones}, K. L., \textit{Supernova remnant candidates from the Parkes 2.4-GHz survey}, 1997, \mnras, 287, 4

\bibitem[{{Morlino} \& {Caprioli}(2012)}]{Morlino2012}
{Morlino}, G. \& {Caprioli}, D., \textit{Strong evidence for hadron acceleration in Tycho's supernova remnant}, 2012, \aap, 538, A81

\bibitem[{{Ahnen} {et~al.}(2017){Ahnen}, {Ansoldi}, {Antonelli}, {Arcaro},
  {Babi{\'c}}, {Banerjee}, {Bangale}, {Barres de Almeida}, {Barrio}, {Becerra
  Gonz{\'a}lez}, {Bednarek}, {Bernardini}, \& {Berti}}]{casa2017}
{Ahnen}, M.~L., {Ansoldi}, S., {Antonelli}, L.~A., {et~al.}, \textit{A cut-off in the TeV gamma-ray spectrum of the SNR Cassiopeia A}, 2017, \mnras, 472,
  2956

\bibitem[{{Zabalza}(2015)}]{Zabalza2015}
{Zabalza}, V., \textit{Naima: a python package for inference of particle distribution properties from nonthermal spectra}, 2015, in International Cosmic Ray Conference, Vol.~34, 34th International Cosmic Ray Conference (ICRC2015), 922

\bibitem[{{H.E.S.S. collaboration}(2022)}]{Kepler2022}
{H.E.S.S. collaboration}, in preparation 
\end{thebibliography}
\end{document}